\documentclass[sigconf]{acmart}

\AtBeginDocument{%
  \providecommand\BibTeX{{%
    \normalfont B\kern-0.5em{\scshape i\kern-0.25em b}\kern-0.8em\TeX}}}
\usepackage{pgfplots} 
\usepackage{hyperref}
\usepackage{enumitem}
\usepackage{stfloats}
\usepackage{xcolor}
\usepackage{url}
\usepackage{soul}
\usepackage{colortbl}
\usepackage{subfigure}
\usepackage[ruled,vlined]{algorithm2e}
\usepackage{caption}
\usepackage{subcaption}
\usepackage{xcolor}
\usepackage{refcount}
\usepackage{float}
\usepackage{siunitx}
\usepackage{booktabs}
\usepackage{graphicx}
\usepackage{multirow}
\usepackage{makecell} 
\usetikzlibrary{tikzmark,calc}
\usepackage{calc}
\usepackage{booktabs}
\usepackage{microtype}
\usepackage{array}
\usepackage{graphicx}
\usepackage{tikz}
\usetikzlibrary{shapes,arrows.meta,positioning,calc,backgrounds}

\SetKwProg{Fn}{Function}{:}{}

\definecolor{vlightgray}{gray}{0.9}

\hypersetup{
 colorlinks=true,   
 citecolor=black,
 urlcolor=blue,
 linkcolor=black,
 bookmarksnumbered=true,     
 bookmarksopen=true,         
 bookmarksopenlevel=1,          
 pdfstartview=Fit,           
 pdfpagemode=UseOutlines,    
 pdfpagelayout=TwoPageRight 
}

\newcolumntype{P}[1]{>{\raggedright\arraybackslash}p{#1}}

\newif\ifdraft
\draftfalse 

\newif\ifrevising
   \revisingfalse 

 \newcommand{\deleted}[1]{{\ifrevising{\relax}\else\relax\fi}}
\usepackage{framed}
\usepackage{xcolor}

\definecolor{TekheletPurple}{HTML}{572F96}

\makeatletter
\providecommand\nopunct{\@addpunct{}}
\makeatother

\title{What's Beyond Copilot?\\
22 AI Systems Developers Want Built}

\title{Looking Beyond Copilot:\\
22 AI Systems Developers Want Built}

\title{You Shall Not Pass!\\Where and Why Developers Draw The Line on AI Autonomy}

\makeatletter
\renewcommand{\@authorsaddresses}{}
\makeatother

\author{%
  Rudrajit Choudhuri$^{1}$ \quad
  Christian Bird$^{3}$ \quad
  Carmen Badea$^{3}$ \quad
  Marco Gerosa$^{2}$ \quad
  Anita Sarma$^{1}$
  \country{}
}
\affiliation{%
  $^{1}$Oregon State University, OR, USA. Email: \{choudhru, anita.sarma\}@oregonstate.edu \\
  $^{2}$Northern Arizona University, AZ, USA. Email: marco.gerosa@nau.edu \\
  $^{3}$Microsoft Research, WA, USA. Email: \{cbird, cabadea\}@microsoft.com%
  \country{}
}

\setcopyright{none}
\acmConference[]{Oregon State University, Northern Arizona University, \& Microsoft Research}{USA}{2026}

\renewcommand{\shortauthors}{Choudhuri et al.}

\begin{abstract}

As AI takes on more software work, the line between human and AI effort is shifting. Where developers draw that line around AI autonomy bears on how we design tools and roles that preserve meaningful work. Drawing on cognitive appraisal theory, work design, and automation research, we conducted a mixed-methods study of 448 professional developers at Microsoft to investigate developers’ accepted levels of AI autonomy across software engineering work.
Most developers accepted AI producing work under their oversight, although accepted autonomy varied substantively across tasks and individuals. Acceptance was lowest for identity-defining, human-facing, and design-oriented work, and higher among developers with more AI experience and risk tolerance. Task accountability was associated with lower odds of allowing AI to \textit{act} on developers’ behalf, whereas task identity was associated with lower odds of granting AI \textit{decision-making} autonomy. Task demands had the opposite effect, increasing willingness to delegate decision-making to AI. Our findings suggest that preferences for AI autonomy reflect how developers cognitively experience their work, highlighting important considerations for designing meaningful work.

\end{abstract}

\begin{document}

\maketitle

\section{Introduction}


Generative AI-powered development tools (hereafter \textit{AI tools}) are transforming how software gets built. AI tools such as GitHub Copilot~\cite{Copilot}, Claude~\cite{Claude}, and in-house assistants today write code, fix bugs, and generate tests, and their capabilities continue to expand. The question is no longer whether AI can do the work; it is \textit{should it, which parts, and how much?} 

This quandary is not new. Long before AI, every wave of automation forced the same reckoning over what to hand to the machine \cite{fitts1951human, de2014fitts, autor2015there}. What is different now is the pace and the reach. AI moves faster and reaches deeper into developers' work, so a single decision to defer can have downstream impacts. If we are not deliberate about this decision and let it settle based on AI tool capabilities, the result can be a set of serious compounding problems ~\cite{miller2025maybe, feng2025gains, storey2026technical, choudhuri2026thinking, afroz2026fast}.

As AI absorbs tasks, developers can miss the work that builds skill and grounds judgment \cite{feng2025charting, storey2026technical, choudhuri2026thinking}, leaving them to ``rubber-stamp'' work they no longer understand well enough to evaluate. Pressure to ship leads them to offload more to keep pace \cite{miller2025maybe}, and as code generation grows cheaper, the same pressure plays out across the pipeline \cite{choudhuri2026copilot}: review, testing, and operations absorb rising volumes of machine-generated work. By the time defects surface, that work has passed through many hands, which makes it costlier to fix \cite{boehm2005software, brooks1987no} and accountability harder to assign \cite{bird2011don, rahman2011ownership}.

Prior work has examined what drives AI adoption~\cite{russo2024navigating, choudhuri2025needs} (e.g., trust~\cite{johnson2023make, choudhuri2025guides, brown2024identifying, wang2023investigating}), which tasks suit automation~\cite{khemka2024toward, lambiase2025exploring}, and where developers want support~\cite{choudhuri2026aiwhereitmatters, pereira2025exploring}, but not where they want control to remain human. 
Without that boundary, we cannot design work, or the tools that structure it, that keeps developers meaningfully engaged. Building on Endsley's levels of automation~\cite{endsley1995outoftheloop}, we distinguish two boundaries that mark it: the \emph{action boundary}, where developers let AI act and produce work artifacts, and the \emph{decision-making boundary}, where they let AI make decisions on their behalf. 

We posit that supporting meaningful work in settings where AI plays a role requires understanding how developers cognitively appraise their work. Drawing on cognitive appraisal theory \cite{lazarus1991emotion, roseman2001appraisal}, work design theory \cite{humphrey2007integrating, lips2009discriminating}, and automation theory \cite{endsley1995outoftheloop}, we examine how developers' task appraisals along dimensions of relevance, identity congruence, accountability, and demands \textit{inform the autonomy they grant AI}. More specifically, we investigate:

\begin{description}
  \item[\textbf{RQ1.}] What level of AI autonomy do developers accept across software engineering tasks, and what predicts where they draw the line?
  \item[\textbf{RQ2.}] What predicts when developers let AI cross the \emph{action} and the \emph{decision-making} boundaries in their work?
\end{description}

We answer these questions through a large-scale survey of 448 professional developers, who reported where they did and did not want AI involved across tasks spanning the software development lifecycle (SDLC).
We classified 1,535 open-ended responses onto a five-level autonomy scale adapted from Endsley and Kiris~\cite{endsley1995outoftheloop}, mapped the accepted AI autonomy levels across SDLC categories, and modeled the accepted level and its two critical transitions (action and decision-making boundaries) against developers' task appraisals and their traits. We interpret these results through the lens of meaningful work design and frame that design as a flight of \textit{Cascading Locks}: a sequence of gates that hold or cede autonomy depending on whether the conditions for meaningful work are met. This framing lets us identify patterns and anti-patterns for meaningful work design.

\section{Related Work}
\label{sec:related}

As AI tools become standard in software engineering (SE) workflows, a growing body of work has examined what drives their adoption~\cite{russo2024navigating, bird2022taking, choudhuri2025needs}. 

Studies based on technology-acceptance models found that workflow compatibility and habitual use are strong drivers of adoption. Trust also emerged as a recurring factor, shaped by tooling capabilities~\cite{johnson2023make, brown2024identifying, wang2023investigating}, individual dispositions~\cite{choudhuri2025needs,choudhuri2026aiwhereitmatters}, and team factors~\cite{cheng2023would, miller2025maybe}. More recent studies moved from overall adoption to task-level differences. Lambiase et al.~\cite{lambiase2025exploring} found higher AI receptivity for artifact-manipulation and information-retrieval tasks and lower receptivity in collaborative contexts. Pereira et al.~\cite{pereira2025exploring} observed stronger adoption for code-intensive work and more limited use in creative aspects. Khemka et al.~\cite{khemka2024toward} reported strong demand for AI in testing, debugging, documentation, and compliance, while Kumar et al.~\cite{kumar2025time} showed that toil-heavy activities such as documentation and environment setup were disproportionately viewed work to minimise and thus strong candidates for AI support. 

Closest to our work, Choudhuri et al.~\cite{choudhuri2026aiwhereitmatters} investigated how developers cognitively appraise different aspects of their work, along dimensions of value, identity, accountability, and demands, to explain where they want or resist AI support across the software lifecycle. That work showed that desired AI involvement was shaped not only by tool capability or trust, but also by what the work meant to developers, and that human oversight remained important even where AI support was welcome. Wanting AI support, however, is not the same as granting AI autonomy. Accepting suggestions, accepting AI-produced artifacts, and allowing AI to act without approval represent different forms of involvement. Work on delegation made this distinction explicit: ceding work to AI is a decision about authority rather than use~\cite{strunk2024delegate}. Ulloa et al.~\cite{ulloa2025product} similarly found that product managers were less willing to hand over work they identified with or felt accountable for. Even though these studies suggest that work appraisals shape delegation decisions, they do not characterize the level of autonomy developers grant once AI starts to be used. 

The \textbf{automation literature} characterizes autonomy as a graded allocation of authority between humans and machines. Endsley and Kiris~\cite{endsley1995outoftheloop} and Parasuraman et al.~\cite{parasuraman2000model} organized automation into levels defined by who decides and who acts, ranging from advisory assistance to full automation. Higher levels have been associated with risks including automation complacency, reduced situation awareness, and out-of-the-loop failures. This perspective underpins machine-in-the-loop design, task-delegability frameworks~\cite{lubars2019ask}, and analyses of work better suited to automation than augmentation~\cite{shao2025future}. Within SE, autonomy has primarily been studied within limited settings. Ghorbani et al.~\cite{ghorbani2023autonomy}, for example, found that developers preferred lower-autonomy tools, although more experienced developers were more receptive to higher-autonomy ones. These framings, however, largely described what tools could do rather than the autonomy developers were willing to grant them. 

Building on these foundations, we investigate where and why developers draw the line on AI autonomy across SE work. We characterize the level of autonomy developers grant AI, examine how those preferences vary across the software lifecycle, and model how developers' cognitive appraisals and individual characteristics predict both accepted AI autonomy and the transitions between autonomy levels.

\section{Theory and Hypotheses}
\label{sec:theory}

When working with machines, humans have to decide how much of the work to keep and how much to cede. 

\textbf{Levels of automation:} The literature treats this division of control between humans and machines as a spectrum~\cite{parasuraman2000model, sheridan1978human, endsley1995outoftheloop}. We adopt the five-level scale from Endsley and Kiris~\cite{endsley1995outoftheloop}, ordered by who decides and who acts, to characterize where developers draw the line on AI autonomy (Table~\ref{tab:levels}). 

\begin{table}[H]
  \centering
  \vspace{-5px}
  \caption{\small The five-level autonomy scale, adapted from Endsley and
           Kiris~\cite{endsley1995outoftheloop}. Each level reflects who decides
           and who acts; AI autonomy increases down the rows.}
    \vspace{-5px}
  \label{tab:levels}
  \footnotesize
  \begin{tabular}{@{}p{0.40\columnwidth}cc@{}}
    \toprule
    \textbf{Level of Automation} & \textbf{Developer} & \textbf{AI} \\
    \midrule
    L1\,\textperiodcentered\,None
      & Decide, Act & \textendash \\
    \addlinespace[2pt]
    L2\,\textperiodcentered\,Decision Support
      & Decide, Act & Suggest \\
    \addlinespace[2pt]
    L3\,\textperiodcentered\,Consensual
      & Decide, Act & Act \\
    \addlinespace[2pt]
    L4\,\textperiodcentered\,Monitored
      & Veto & Decide, Act \\
    \addlinespace[2pt]
    L5\,\textperiodcentered\,Full Automation
      & \textendash & Decide, Act \\
    \bottomrule
  \end{tabular}
\vspace{-5px}
\end{table}

We chose it over frameworks that decompose automation across information-processing stages, such as Parasuraman et al.~\cite{parasuraman2000model}, which assigns separate levels to acquisition, analysis, decision, and action. A single ordered scale gives one level of human–AI control per task, which keeps levels comparable across the heterogeneous tasks in SE work.

Table~\ref{tab:levels} shows these five levels, ordered by how much AI decides and acts without the developer. 
At L1, the developer decides and acts, and AI has no role. 
At L2, AI suggests or flags while the developer decides and produces the work. 
At L3, AI may act, but the result takes effect only after the developer signs off.
At L4, AI decides and acts by default, and the developer retains an optional veto. At L5, AI decides and acts on its own with minimal human involvement. 

The two transitions on this scale that impact work design revolve around the kind of control a developer cedes. At the \emph{action boundary} (L2$\to$L3), AI moves from advising to producing the artifact or carrying out the action. At the \emph{decision-making boundary} (L3$\to$L4), the developer's sign-off moves from required to optional. 

\textbf{Hypotheses: Predictors of levels of automation:}
The goal of our work is to investigate where and why developers cede autonomy to AI and to what extent. While trust is an obvious precursor to whether developers use AI, it concerns the \emph{``can AI do it''} question, which prior work has addressed~\cite{lee2004trust, choudhuri2025guides, wang2023investigating, johnson2023make}. In this paper, we investigate how developers answer the \emph{``should AI do it and to what extent''} question.

Humans are meaning-makers; we seek significance and value in our work~\cite{lips2009discriminating}. At work, we implicitly evaluate a task by asking: Is this important to me? Does it align with what I want to do? Am I responsible if it fails? Can I handle its demands? \textbf{Cognitive appraisal theory}~\cite{lazarus1991emotion, roseman2001appraisal} formalizes these judgments across dimensions of importance, congruence with one's motivations or identity, accountability, and cognitive demand. These appraisals shape how people cope with work~\cite{campbell2020cognitive} and predict engagement, persistence, and discretionary effort~\cite{meyer1991three}. Work-design research \cite{humphrey2007integrating} adds that motivational, social, and contextual job characteristics explain much of the variance in work engagement and productivity. 

Choudhuri et al.~\cite{choudhuri2026aiwhereitmatters} showed that these appraisals predict developers' openness to and use of AI. Building on this, we use the appraisals to predict how much autonomy developers cede to AI in SE tasks. Specifically, we study four appraisals: \textit{Value} and \textit{Identity} are motivational appraisals~\cite{lips2009discriminating}, \textit{Accountability} a social one~\cite{tetlock1983accountability}, and \textit{Demand} a contextual one~\cite{bakker2007job}.

\textbf{Task Value} is the perceived importance of a task to project success and to personal goals~\cite{hackman1976motivation}. Valuable work raises both the benefit of finishing it faster and the cost of getting it wrong; developers treat the work that matters most as the work they most want to keep under their own control.
\textbf{H1.} \textit{Higher task value lowers the AI autonomy developers accept}. We expect developers to retain autonomy for tasks they consider valuable.

\textbf{Task Identity} is the degree to which a developer identifies with a task and enjoys it for its own sake~\cite{ryan2000self}. Identity-defining tasks confer ownership and a sense of craft~\cite{kahn1990psychological, koestner2002attaining}, which we hypothesize developers protect by keeping them under their control.
\textbf{H2.} \textit{Higher task identity lowers the AI autonomy developers accept}.

\textbf{Task Accountability} is the perceived responsibility a developer feels for a task's outcome~\cite{hall2017accountability}. Anticipated answerability raises vigilance and the wish to oversee the work before signing off as their own~\cite{tetlock1983accountability, lerner1999accounting}. The more autonomy they grant AI, the less they can check, so we hypothesize:
\textbf{H3.} \textit{Higher task accountability lowers the AI autonomy developers accept}.

\textbf{Task Demand} is the cognitive effort a task imposes~\cite{bakker2007job}. Demanding work strains a developer's resources and increases the likelihood of offloading work \cite{bakker2007job, feng2025gains}, so developers should give AI higher autonomy where demand is high.
\textbf{H4.} \textit{Higher task demand raises the AI autonomy developers accept}.

\textbf{Task type.} Appraisals are tied to the type of work one performs. For example, human-facing work such as mentoring, designing, and stakeholder communication carries the meaning, judgment, and relationships that make work feel one's own~\cite{humphrey2007integrating, lips2009discriminating}, whereas well-scoped or rote system tasks such as test generation, environment setup, and documentation carry less of it. Where a task lies along that range should relate to how much autonomy a developer is willing to give AI. We therefore expect the autonomy developers accept to differ across the SDLC categories.
\textbf{H5.} \textit{The AI autonomy developers accept differs across task types}.

\textbf{Crossing the boundaries.} Letting AI draft work developers sign off on (action boundary) and letting AI act unless developers stop it (decision-making boundary) are very different levels of autonomy, as defined in Table~\ref{tab:levels}. Both are conditions that tie to meaningful work design: people experience work as meaningful when they author it and when they hold the decisions that govern it, and that sense of ownership is what sustains engagement and intrinsic motivation~\cite{humphrey2007integrating, lips2009discriminating, kahn1990psychological, ryan2000self}. Each boundary removes one of these conditions, so we model them separately and expect the appraisals to act on each as they do on the overall level.
\textbf{H6.} \textit{Higher task value, identity, and accountability lower the odds that developers cross each boundary; higher task demand raises the odds}.

\textbf{Controls.} We control for developers' SE and AI experience, since both shape attitudes toward AI~\cite{butler2025dear, crowston2025deskilling}. Experience can calibrate these judgments: time spent building software and time spent working with AI tools both temper what a developer expects AI to handle and how readily they delegate to it. Individual traits can also condition how appraisals translate into the autonomy a developer accepts. Prior work has shown that individuals' risk tolerance and technophilic motivations~\cite{choudhuri2025guides, anderson2022measuring} are associated with stronger AI-adoption dispositions. 
We expect these traits and experience measures to moderate the hypothesized appraisal effects.

\section{Method}
\label{sec:method}

Our goal was to investigate where and why developers drew the line on AI autonomy across their daily SE tasks. We studied professional software developers across Microsoft, an AI-forward technology company employing more than 50,000 developers worldwide, spanning a diverse set of products, teams, roles, processes, and stakeholder contexts. The scale, combined with participants' exposure to both emerging and mature AI tools, makes it a rich setting for our study.

\subsection{Study Design}
\label{sec:design}

We surveyed developers about how they cognitively appraised different aspects of their daily work and where they did or did not want AI involved, following the appraisal-survey design of~\cite{choudhuri2026aiwhereitmatters}. We classified the open-ended responses onto a five-level autonomy scale (Section~\ref{sec:analysis}) and modeled the result against task appraisals and developer traits, characterizing what set the autonomy level developers accepted and the transitions where they granted AI \textit{action} and \textit{decision-making} authority. Our study was approved by the company's IRB.

\subsubsection{\textbf{Survey instrument design}}
\label{sec:instrument}

We followed Kitchenham's survey guidelines~\cite{kitchenham2008personal} and drew on established theoretical frameworks and validated scales (Table~\ref{tab:constructs}) to capture how developers appraise their daily work and how much autonomy they would grant AI on it. To represent that work, we adopted the grounded taxonomy of software engineering (SE) tasks across five SDLC categories (Table~\ref{tab:taxonomy}) from~\cite{choudhuri2026aiwhereitmatters}. We refined the survey instrument through iterative pilots with researchers and developers. The survey was structured as follows.

\begin{table}[!ht]
\footnotesize
\caption{\small Theoretical constructs and instruments~\cite{choudhuri2026aiwhereitmatters}}
\vspace{-5px}
\label{tab:constructs}
\centering
\begin{tabular}{>{\raggedright\arraybackslash}m{3.5cm} >{\raggedright\arraybackslash}m{4.5cm}}
\hline
\textbf{Construct} & \textbf{Instrument} \\
\midrule
Value & Job Characteristics Model~\cite{fried1987validity, trinkenreich2024predicting} \\ 
Identity & Self-Determination Theory~\cite{ryan2000self} \\ 
Accountability & Felt Accountability Scale~\cite{hall2017accountability} \\ 
Demands & Job Demands-Resource Model~\cite{bakker2007job}\\
\midrule
Levels of Automation & Endsley \& Kiris Framework \cite{endsley1995outoftheloop} \\ 
Risk Tolerance, Technophilia & Cognitive Style Facet Survey \cite{choudhuri2026aiwhereitmatters, choudhuri2025guides} \\ 

\hline
\end{tabular}
\end{table}

\vspace{-2mm}
After providing informed consent, developers reported their SE and AI tool experience and their dispositions toward AI (risk tolerance and technophilia), then selected the 2--3 categories from Table~\ref{tab:taxonomy} that best reflected their work. To reduce fatigue, the meta-work category (applicable to all developers) was not a default option and appeared only for participants who selected two other categories, ensuring that no participant answered more than three category blocks.

Within each category block, participants rated how they appraised each task along four dimensions: value, identity, accountability, and demands~\cite{choudhuri2026aiwhereitmatters}, which served as our model predictors. We used single-item measures to keep the survey tractable and reduce participant fatigue, as these retain psychometric validity for well-scoped constructs~\cite{matthews2022normalizing}.

\begin{table}[h]
\centering
\caption{\small Grounded SE task taxonomy from~\cite{choudhuri2026aiwhereitmatters}.}
\label{tab:taxonomy}
\vspace{-5px}
\footnotesize
\begin{tabular}{@{}p{0.25\columnwidth} >{\raggedright\arraybackslash}p{0.65\columnwidth}@{}}
\toprule
\textbf{SDLC Category} & \textbf{Tasks} \\
\midrule
Development
& Coding, Bug fixing, Performance optimization, Refactoring, AI integration \\
Design \& Planning
& System design, Requirements engineering, Project planning \& management \\
Quality \& Risk
& Testing/QA, Code review, Security \& compliance \\
Infrastructure \& Ops
& DevOps (CI/CD), Environment setup \& maintenance, Infrastructure monitoring, Customer support \\
Meta-work
& Learning, Research, Documentation, Stakeholder communication, Mentoring \\
\bottomrule
\end{tabular}
\vspace{-2mm}
\end{table}

Each block then posed two open-ended questions about where developers wanted and did not want AI: “\textit{Where do you want AI to play the biggest role for your [category]-heavy activities over the next 1--3 years?}” and “\textit{What aspects of your [category]-heavy activities do you not want AI to handle and why?}” 
The autonomy preferences are contingent on task type and contextual constraints, and the open-ended questions let preferences emerge naturally. 
We subsequently classified responses onto the five-level autonomy scale (Section~\ref{sec:theory}).

We administered the survey in Qualtrics~\cite{qualtrics2025}. Closed-ended items used a 5-point Likert scale with a sixth ``I'm not sure'' option. The survey took 10--12 minutes to complete. To support data quality and reduce response bias, we included attention checks and randomized question order within blocks. 

We refined the instrument in pilot rounds with 40 developers, checking clarity and task coverage. These pilot responses were excluded from the analysis. The survey instrument is in~\cite{supplemental}.

\subsubsection{\textbf{Data collection}}
\label{sec:survey}

We emailed the survey to a random-stratified sample of 8,000 developers across varied roles, teams, and regions, with one follow-up a week later. Participation was voluntary and anonymous. The survey returned 1,193 responses, a 14.9\% response rate consistent with prior large-scale SE surveys~\cite{storey2019towards, choudhuri2025needs}. We dropped incomplete ($n{=}152$), straight-lined or otherwise patterned responses ($n{=}59$), those that failed an attention check ($n{=}98$), and ones from developers reporting no AI experience ($n{=}24$). 

The autonomy classification depended on the open-ended responses, so we also dropped participants whose responses were unsubstantive ($n{=}412$). A field counted as substantive when it held content beyond an empty string or a set of filler tokens (e.g., ``n/a,'' ``yes,'' ``no,'' ``x,'' ``nil,'' ``nope,'' ``idk,'' or ``no comment''). One researcher applied this rule to every open-ended field. The remaining 448 developers provided 1,535 task-responses (we describe the decomposition in Section~\ref{sec:coding}), of which the 1,476 with complete predictor data feed the regression models in Section~\ref{sec:results}. Most respondents were based in North America and self-identified as men, spanning a wide range of SE and AI experience. Demographics are in~\cite{supplemental}.

\subsection{Data Analysis}
\label{sec:analysis}

Our analysis proceeded in two steps. We first classified each open-text task response onto the five-level autonomy scale (Table~\ref{tab:levels}), then modeled the classified responses against task appraisals and developer traits (Section~\ref{sec:theory}). Then, we conducted reflexive thematic analysis of the same responses to characterize which aspects of work developers wanted to retain at each autonomy level. 

\subsubsection{\textbf{Autonomy level classification}}
\label{sec:coding}

Each open-text response set recorded, in participants' own words, the role they
wanted AI to play in a task-category (\textsc{Want}) and the role they refused to
grant it (\textsc{Resist}). The pair defines the ceiling, the highest level of autonomy a participant tolerated before \textsc{Resist} blocked more. 
To classify the responses, we used a rule-based algorithm (Algorithm~\ref{alg:coding}) that classified each response to an autonomy level. When the text was ambiguous between two adjacent levels, fuzzy labeling assigned a boundary level rather than forcing one. The algorithm comprised the following components:

\setlength{\textfloatsep}{6pt plus 2pt minus 5pt}
\begin{algorithm}[t]
\DontPrintSemicolon
\footnotesize
\SetKwInOut{Input}{Input}\SetKwInOut{Output}{Output}
\SetKwFunction{classify}{classify}
\Input{\textsc{Want} field $w$ and \textsc{Resist} field $r$ for one
(participant, category) cell}
\Output{One or more rows $\langle\text{task},\,\text{source},\,\text{label},\,\text{rationale}\rangle$}
\BlankLine

\Fn{\classify{$s$}}{
\tcp{walk string $s$; stop at first match (see marker table in~\cite{supplemental})}
\uIf{REFUSAL of all AI for the task, no carve-out}{\Return \textbf{L1}}
\uElseIf{REFUSAL $+$ SUGGESTION in limited cases}{\Return \textbf{L1--L2}}
\uElseIf{AI contributes SUGGESTION; takes no ACTION or produces no ARTIFACT}{\Return \textbf{L2}}
\uElseIf{AI produces an ARTIFACT the human is framed as extending}{\Return \textbf{L2--L3}}
\uElseIf{APPROVAL required \emph{before} AI proceeds with an ACTION}{\Return \textbf{L3}}
\uElseIf{AI ACTION language present $+$ no APPROVAL/VETO anchor}{\Return \textbf{L3--L4}}
\uElseIf{AI ACTION by default; human keeps VETO}{\Return \textbf{L4}}
\uElseIf{AI ACTION with MONITORING-ONLY}{\Return \textbf{L4--L5}}
\uElseIf{AI ACTION with NO-HUMAN-CONTROL}{\Return \textbf{L5}}
}
\BlankLine

$W \leftarrow \textsc{Tasks}(w)$;\ \ $R \leftarrow \textsc{Tasks}(r)$
\tcp*{ see hyponym table in~\cite{supplemental}; split clauses with distinct nouns}
pair $W$ and $R$ by the same-task test;
set $\textit{source}\in\{\textsc{Want},\textsc{Resist},\textsc{Both}\}$ per row\;
\BlankLine
$s \leftarrow w$ if \textsc{Want}, $r$ if \textsc{Resist}, $w \Vert r$ if \textsc{Both}\tcp*{source}
$\textit{label} \leftarrow \classify{s}$\;
\Return the coded row(s) with rationale\;
\caption{\small Autonomy-level classification of a (participant, category) response. \textsc{Tasks}($\cdot$) splits each field into per-task rows; \protect\classify\ returns the corresponding autonomy-level.}
\label{alg:coding}
\end{algorithm}

\noindent\textbf{Marker and hyponym tables.} The algorithm used two tables in the
classification process (see~\cite{supplemental}). The hyponym table mapped surface phrases onto canonical tasks (e.g., \emph{write code}, \emph{programming}, \emph{code-related} all resolved to the Coding task), which the algorithm then used to pair \textsc{Want} and \textsc{Resist} responses when both named the same task and split them otherwise. The marker table was a set of autonomy features, each a closed set of case-insensitive substrings: \textsc{Refusal} (AI does no part of the task), \textsc{Suggestion} (AI suggests), \textsc{Action} (AI acts), \textsc{Artifact} (AI produces an artifact), \textsc{Approval} (the human signs off before AI's work takes effect), \textsc{Veto} (AI decides and acts, the human can override afterward if needed), \textsc{Monitoring-only} (AI acts, the human only watches), and \textsc{No-human-control} (AI acts without notable human involvement). We synthesized both tables from the pilot responses, then finalized them against the full response set to capture forms the pilot missed. To guard against circularity, the external coder (described below) saw neither table during development.

\noindent\textbf{Unit of analysis.} The unit of analysis was the triple (participant, category, task). Participants often described several tasks within a single response set (per category) and named different tasks across the two fields. \textsc{Tasks}($\cdot$) therefore separated each field into one row per task using the hyponym table. We then paired \textsc{Want} and \textsc{Resist} when both named the same task and split them otherwise, setting each row's source to \textsc{Want}, \textsc{Resist}, or \textsc{Both}. This analysis yielded 1{,}535 task-responses.

\noindent\textbf{Level assignment.} For each row, the \textsc{classify()} subroutine applied a fixed sequence of binary tests ordered by increasing AI autonomy, stopping at the first marker that fired (see Algorithm~\ref{alg:coding}). The tests capped each row at the highest autonomy the participant tolerated: any refusal or approval requirement was reached before higher-autonomy markers. 
Around 8\% of rows received a boundary label (e.g., L2-L3, L3-L4). For the descriptive distribution (Figure~\ref{fig:autonomy_matrix}), we conservatively collapsed each to its lower adjacent level (full distribution in~\cite{supplemental}). The regression models instead entered each as an interval-censored observation between its two levels, preserving its encoded uncertainty.

\noindent\textbf{AI-council-based classification.} We ran algorithm~\ref{alg:coding} with an AI-council comprising an
ensemble of three frontier Large Language Models from distinct families (\texttt{gpt-5.4},
\texttt{gemini-3-flash}, \texttt{claude-sonnet-4.6}; high-reasoning mode~\cite{dunivin2024scalable, hila2025deductive}). We selected the three from different providers to
reduce model-specific blind spots, inductive biases, and failure modes~\cite{alshaikh2026prompt, li2026deductive, ashwin2026using}. Agreement across model families provided stronger convergent evidence than agreement within one, where shared data and optimization objectives could produce shared errors~\cite{wu2026council}. No model saw the others' outputs. We reconciled the three label sets by majority vote: a label held when at least two coders agreed.

\noindent\textbf{Human validation and IRR.} To verify the classification, two
researchers independently re-coded a random sample of 67 participant-response sets from the 448 participant pool ($\pm10\%$ margin at 95\% confidence). One researcher was not involved in developing the algorithm and acted as an external check against circularity. We measured inter-rater reliability (IRR) with Krippendorff's $\alpha$ complemented by pairwise quadratic-weighted Cohen's $\kappa$ and three-rater percent agreement~\cite{gwet2014handbook}. Researcher agreement was $\kappa = 0.94$, human--council agreement $\kappa = 0.95$ and $0.92$, and overall three-rater agreement $\alpha = 0.93$, indicating strong agreement for the analyses that follow.

\vspace{-3mm}
\subsubsection{\textbf{Mixed-methods analysis}}
\label{sec:mma}

We modeled the coded task-responses against participants' cognitive appraisals and individual characteristics (recall Section~\ref{sec:theory}). Because the design involved repeated measures within participants and across tasks, we used mixed-effects regression~\cite{gelman2007data}. For RQ1, we fit a Cumulative Link Mixed Model (CLMM) on the five-level ordinal scale against z-standardized predictors, accommodating the four fuzzy boundary labels (e.g., L2--L3) as interval-censored observations~\cite{hedeker1994random, xie2000random}. For RQ2, we fit a logistic Generalized Linear Mixed Model (GLMM) at each of the two autonomy boundary transitions (action: L2$\to$L3, decision-making: L3$\to$L4), modeling whether a participant's accepted autonomy crossed that boundary. All Variance Inflation Factors (VIFs) were $<2$~\cite{hair2009multivariate}, indicating no multicollinearity concerns. Model specifications appear in Section~\ref{sec:results}.

To understand how participants reasoned about levels of AI autonomy across aspects of their work, we conducted reflexive thematic analysis~\cite{braun2006using, braun2022conceptual} of the classified open-ended responses. We inductively coded participants' explanations and developed themes characterizing the work aspects tied to each autonomy level. Two researchers iteratively refined the codes and themes, with the research team resolving disagreements through negotiated agreement. The resulting themes organize Figure~\ref{fig:autonomy_matrix} and explain participants' autonomy preferences in Section~\ref{sec:results}. Participants are referenced as P1--P448 hereafter.

\subsection{Threats to Validity}
\label{sec:threats}

\noindent\textbf{Construct validity.} We measured constructs with self-reported items grounded in established theory. Still, surveys can introduce bias or misinterpretation. We mitigated this by involving practitioners in survey design, piloting, randomizing blocks, adding attention checks, and screening patterned responses. We elicited autonomy preferences as open-ended text rather than fixed ratings, since the level a developer deems appropriate is contingent on contextual constraints that cannot be reliably assessed psychometrically. Finally, to mitigate forced categorization, we introduced fuzzy labels for boundary-ambiguous responses, treating them as interval-censored data in the primary model, and confirmed that the conclusions were robust under alternative operationalizations (Section~\ref{sec:rq1}).

\noindent\textbf{Internal validity.} As a cross-sectional study~\cite{stol2018abc}, we report associations, not causation. Self-selection is possible, since developers with stronger views about AI may be more likely to respond. As with all survey-based work, our results reflect self-reported perceptions. To guard against classifier bias, we ran an AI council of three models on a fixed, rule-based algorithm, reconciled by majority vote~\cite{wu2026council, alshaikh2026prompt}, with an external recoder checking against circularity, with strong agreement throughout (Section~\ref{sec:coding}). Still, inferring a level from brief free text is imperfect. We strengthened validity by triangulating the quantitative results with qualitative data and aligning with theory. The company's pro-AI culture, if anything, biases against us, making the boundaries developers still refuse a conservative read of where they want to retain autonomy.

\noindent\textbf{External validity.} All participants were from one AI-forward multinational organization, spanning diverse teams, roles, geographies, domains, and processes. This limits generalizability to smaller organizations or open-source contexts. Despite this limitation, we argue our findings are of value to the broader community, because it has been shown that single-case studies make meaningful contributions to scientific discovery~\cite{flyvbjerg2006five}  
that complement research on broader populations~\cite{basili02}.

\section{Results}
\label{sec:results}
Section~\ref{sec:rq1} presents where and why participants drew the line on AI autonomy, followed by Section~\ref{sec:rq2} that presents what predicted whether participants crossed the action boundary (L2$\to$L3) and the decision-making boundary (L3$\to$L4).

\subsection{RQ1: Accepted Autonomy Levels and their Predictors}
\label{sec:rq1}
\begin{figure*}[t]
  \centering
  \includegraphics[trim=50 20 10 0, clip, width=\textwidth]{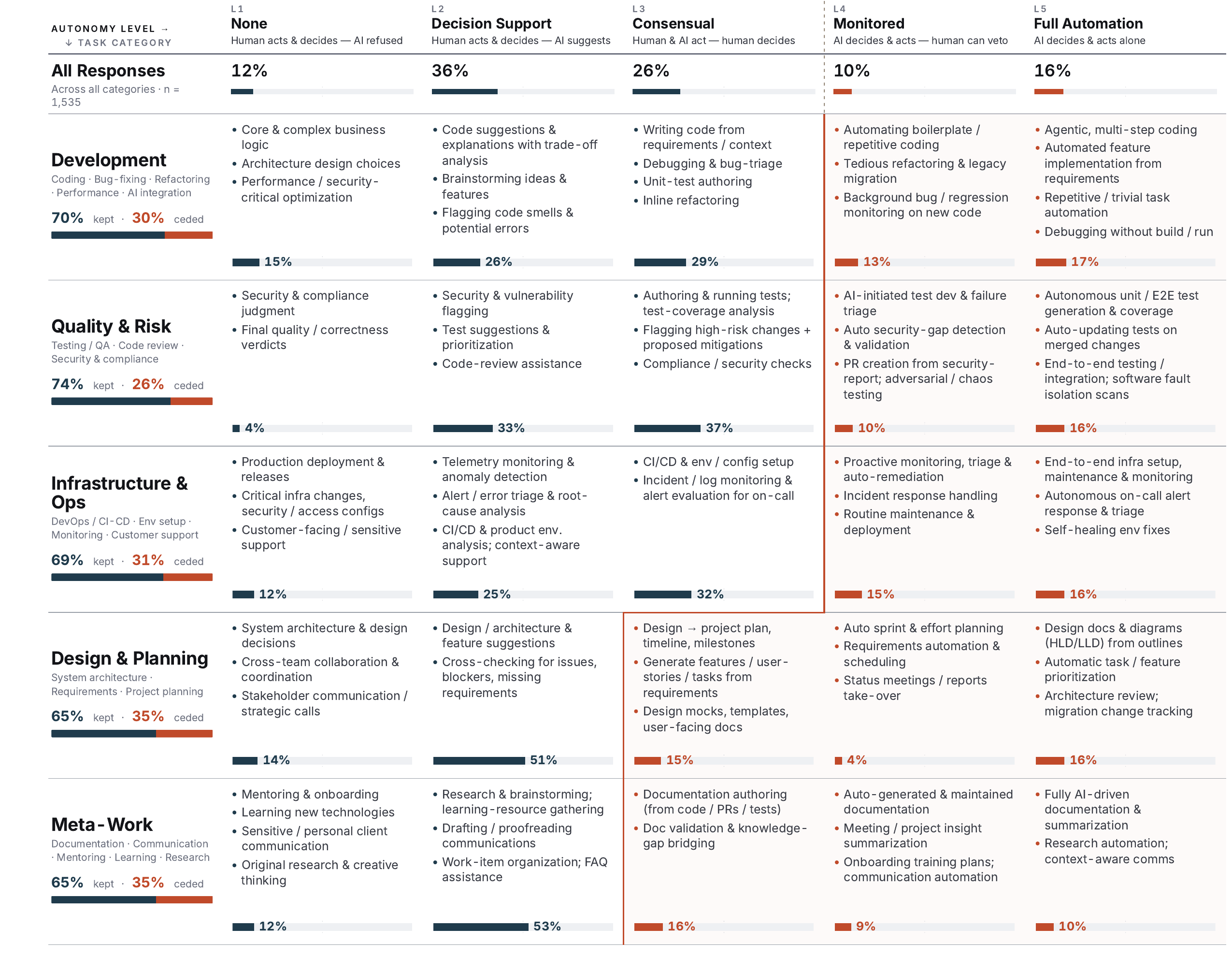}
  \vspace{-5px}
  \caption{\small Participant response distribution ($n{=}1{,}535$ responses; 448~developers) across the five autonomy levels. Rows group responses by SDLC category, with the top row pooling all categories; columns order the levels L1 to L5, defined by who decides and who acts. Each row reports the share of responses at each level, and the proportion of autonomy participants kept versus ceded to AI, split at each category's median accepted level; each cell lists representative aspects of work mentioned for that category and level.
  }
\vspace{-4mm}
  \label{fig:autonomy_matrix}
\end{figure*}

We classified all 1{,}535 participant responses on the five-level Endsley--Kiris autonomy scale (Table~\ref{tab:levels}), coding each response as the highest autonomy a participant accepted for the task (L1 = no AI involvement through L5 = full automation). Figure~\ref{fig:autonomy_matrix} places each SDLC category across the five levels, reporting the share of responses at each level, the proportion of autonomy participants kept versus ceded, and representative aspects of work named at each cell.

\subsubsection{Autonomy Level Ceilings}\textbf{Finding 1: Most participants accepted AI-produced work without ceding decision-making to AI ($\le$L3).}  
The median accepted level was L3, where AI produces the artifact and it takes effect only after the developer's explicit approval. 74\% of responses fell at or below L3, whereas 26\% let AI decide and proceed without required approval (L4, optional veto: 10\%; L5, full automation: 16\%). Participants accepted AI across a broad range of work, including inline refactoring, authoring tests, CI/CD pipeline setup, creating user stories from requirements, and authoring docs from PRs (Figure~\ref{fig:autonomy_matrix}, $\le$L3). As P219 put it: \textit{``I'd love for [AI] to generate the code, but in a way that makes it easy for me to follow along and correct it''} (P219, L3). 

Participants wanting to retain oversight is intuitive and consistent with HCI work on user control over intelligent assistive systems~\cite{limerick2014experience} and a preference for human supervision of automation~\cite{bainbridge1983ironies, lubars2019ask}. We now turn to the nuance: where and why participants placed that oversight, and how it varied with the task type, participants' cognitive appraisals, and individual dispositions towards AI support in work.

\textbf{Finding 2: In system-facing work, most participants delegated bounded execution and verification tasks, while retaining oversight on context-sensitive judgments (L3).} The three system-facing categories (Development, Quality \& Risk, and Infrastructure \& Ops) all had a median accepted autonomy level of L3 (Figure~\ref{fig:autonomy_matrix}): most participants positioned AI primarily as a collaborator operating under their approval. As P106 explained, \textit{``I would like to remain a code-reviewer for my AI-generated code and basically steer it in the right direction''} (P106, L3).
 
Within the Quality \& Risk category (26\% of responses~$>$L3), participants were most willing to delegate work they characterized as routine verification or operational toil. They accepted high levels of autonomy ($\ge$L3) for automatically generating tests, detecting security risks, configuring adversarial or chaos-testing environments, creating pull requests from security reports, and performing security and compliance scans. One participant envisioned \textit{``more automated tests built by AI that even a PM could specify what they want to test and AI handles it''} (P36, L5). In contrast, participants insisted on retaining oversight and approval authority for activities requiring contextual evaluation, such as code review, pre-merge security assessment, and test-selection decisions. P196 noted, \textit{``I would be happy for AI to take over code review if it's capable. Testing is the same. Both would require mandatory human post-review''} (P196, L3). The distinction was not necessarily between technical and non-technical work. Rather, participants differentiated between tasks they viewed as automatable toil and those requiring interpretation, prioritization, or judgment to guard against complacency and automation bias~\cite{parasuraman2010complacency}.

\textbf{Finding 3: In design and human-facing work, most participants limited AI to decision support (L2), retaining action and decision-making.} These categories had a median accepted level of L2: participants welcomed AI suggestions but wanted to remain the primary actor and decision-maker, citing concerns about AI's judgment, vision, empathy, and accountability. P276 explained, \textit{``I don't want AI to handle final judgment calls on ambiguous or high-stakes decisions, because these often require human intuition, contextual awareness, and accountability that AI can't fully replicate or own''} (P276, L2). This preference was strongest for activities participants viewed as inherently human. Within Meta-work, mentoring, sensitive client communication, learning, and creative thinking were often retained entirely at L1: \textit{``Mentoring and onboarding shouldn't be done by an AI. It's a human thing that shouldn't be handed off to technology''} (P101, L1). Participants described these activities as opportunities to develop expertise, cultivate others, and exercise interpersonal judgment.

At the same time, they readily accepted AI assistance for drafting and proofreading communications, organizing work items, gathering learning resources, and brainstorming (L2). P125, for example, described using AI to support learning: \textit{``when learning new technologies, I generally have very specific questions that I might feel embarrassed to ask a more knowledgeable person...and get an [AI] expert and specific response back is transformative in the learning process''} (P125, L2). Documentation was delegated further (L3), reflecting more comfort with AI codifying and organizing existing knowledge than with decisions that shape people, strategy, or growth.

\subsubsection{Predictors of accepted AI autonomy}
The descriptive analysis identified where participants drew the line on AI autonomy; we next examined what factors predicted that line and why participants facing the same task diverged. To investigate this, we fit a cumulative-link mixed model (CLMM) with a logit link~\cite{christensen2023ordinal} over the $n{=}1{,}476$ (of 1{,}535) task responses with complete predictor data. The model included z-standardized appraisal and trait predictors (recall Table~\ref{tab:constructs}), SDLC category as a fixed effect (Development as the baseline), and a participant-level random intercept to account for repeated measures.
The ordinal outcome was the accepted autonomy level, with the four fuzzy boundary labels treated as interval-censored observations~\cite{xie2000random}. All predictors were estimated within a single model. For appraisal and trait measures, we report standardized latent-scale coefficients ($\beta$), representing the change in accepted autonomy associated with a one-standard-deviation increase in the predictor. For SDLC categories, we report odds ratios (OR) relative to Development (baseline). Model fit is summarized using marginal and conditional $R^2$~\cite{nakagawa2013r2,mckelvey1975statistical}. Results are in Table~\ref{tab:clmm}.

\begin{table}[t]
\centering
\caption{\small Predictors of accepted AI autonomy (CLMM; $n{=}1{,}476$ task-level responses from 448 developers). Appraisals and traits: standardized $\beta$ (95\% CI) with effect sizes ($f^2$). SDLC categories: odds ratios vs.\ Development (${>}$1 more, ${<}$1 less). Shaded rows are statistically significant ($^{*}p{<}$.05, $^{**}p{<}$.01, $^{***}p{<}$.001, BH-corrected~\cite{thissen2002quick}). We consider Std. effect ($r$)${<}0.02$ no effect, $r\in[0.02,0.15)$ small, $r\in[0.15,0.35)$ medium, $r{>}0.35$ large~\cite{cohen2013statistical}.}

\label{tab:clmm}
\setlength{\tabcolsep}{4pt}
\renewcommand{\arraystretch}{1.15}
\definecolor{sigshade}{gray}{0.90}
\begin{tabular}{@{}lcccc@{}}
\toprule
Predictor & $\beta$ & 95\% CI & $r$ & $p$ \\
\midrule
\multicolumn{5}{@{}l}{\textit{Task appraisals}} \\
Task Value          & 0.06  & [-0.07, 0.18]  & 0.03  & 0.373 \\
\rowcolor{sigshade}
\textbf{Task Identity} & \textbf{-0.13} & \textbf{[-0.24, -0.02]} & \textbf{0.07} & \textbf{0.017*} \\
Task Accountability & -0.07 & [-0.19, 0.05]  & 0.04 & 0.276 \\
Task Demand         & 0.05  & [-0.06, 0.17]  & 0.03  & 0.373 \\
\addlinespace[2pt]
\multicolumn{5}{@{}l}{\textit{Developer traits}} \\
SE Experience       & -0.10 & [-0.33, 0.13]  & 0.06 & 0.393 \\
\rowcolor{sigshade}
\textbf{AI Experience}  & \textbf{0.26} & \textbf{[0.07, 0.45]} & \textbf{0.14} & \textbf{0.007**} \\
\rowcolor{sigshade}
\textbf{Risk Tolerance} & \textbf{0.33} & \textbf{[0.07, 0.60]} & \textbf{0.18} & \textbf{0.013*} \\
Technophilia        & 0.02  & [-0.22, 0.25]  & 0.01  & 0.898 \\
\midrule
\multicolumn{5}{@{}l}{\textit{SDLC categories vs. Development (baseline)}} \\
 & OR & 95\% CI & $r$ & $p$ \\
 \midrule
\rowcolor{sigshade}
\textbf{Design \& Planning} & \textbf{0.42} & \textbf{[0.31, 0.57]} & \textbf{0.48} & $<$$\textbf{.001***}$ \\
\rowcolor{sigshade}
\textbf{Quality \& Risk}    & \textbf{1.67} & \textbf{[1.18, 2.38]} & \textbf{0.28}  & \textbf{0.004**} \\
Infrastructure \& Ops       & 0.79          & [0.58, 1.19]          & 0.13          & 0.053 \\
\rowcolor{sigshade}
\textbf{Meta-work}          & \textbf{0.33} & \textbf{[0.24, 0.46]} & \textbf{0.61} & $<$$\textbf{.001***}$ \\
\midrule
\multicolumn{2}{@{}l}{$R^2_m$, $R^2_c$} & \multicolumn{3}{l}{0.07, 0.49} \\
\bottomrule
\end{tabular}
\end{table}

\textbf{Finding 4: Among the four appraisals, only Task Identity predicted accepted AI autonomy, and negatively.} Task Identity was negatively associated with accepted autonomy ($\beta{=}{-}0.13$, $p{=}0.017$), \textbf{supporting H2}: the more participants identified with a task, the less autonomy they ceded to AI. Task Value, Accountability, and Demand were not significant predictors (no support for H1, H3, or H4). Participants retained autonomy over the tasks that defined them as professionals even when they accepted that AI could do it: \textit{``I do not want to hand AI a design and have it write all the code because I enjoy coding and the challenges of solving problems in programming.''} (P176, L2). This finding aligns with self-determination theory, under which the activities people pursue for their own sake are the ones that satisfy the needs for autonomy and competence~\cite{ryan2000self}, aspects contributing to one's sense of craft, source of achievement, and self worth~\cite{kahn1990psychological, koestner2002attaining}. 


\textbf{Finding 5: Participants' AI experience and Risk Tolerance were positively associated with their accepted AI autonomy level.} AI Experience ($\beta{=}0.26$, $p{=}0.007$) and Risk Tolerance ($\beta{=}0.33$, $p{=}0.013$) were both associated with higher accepted autonomy. Participants with more experience using AI tools and those more comfortable with uncertainty were more willing to let AI act with less oversight. As P288 (``very experienced'' with AI and reporting high risk tolerance) put it: \textit{``Development in general should be taken over by AI. If the human can correctly word out what needs to be done, the actual implementation shouldn't require too much oversight''} (P288, L5). Neither SE Experience nor Technophilia predicted the accepted level: years of building software did not by itself predict ceding autonomy to AI, nor love for new technology. 

\textbf{Finding 6: Task type predicted the participants’ accepted autonomy level.} 
Task type made a difference. Relative to Development (baseline), participants accepted lower autonomy for Meta-work (OR$=0.33$, $p<0.001$) and Design \& Planning (OR$=0.42$, $p<0.001$). Meta-work (mentoring, communication, onboarding, and learning) had the strongest resistance. Participants preferred to retain control over activities requiring interpersonal judgment, intent, and relationship management: \textit{``I don't want [AI] to help mentor people. Relationships are important''} (P85, L1). Another explained, \textit{``I want AI suggestions for communication with external partners...I don't want wholesale retranslation
''} (P182, L2). Participants also limited autonomy in Design \& Planning, viewing these activities as highly consequential and costly to reverse. P10 remarked, \textit{``AI can support but shouldn't lead, as misalignment here can derail entire projects''} (P10, L2). 

In contrast, Quality \& Risk had higher odds of accepted autonomy than Development (OR$=1.67$, $p=0.004$). Participants were more willing to delegate testing and review activities, often framing AI as an additional safeguard for identifying \textit{``bugs, regressions, [performance] bottlenecks, and potential security issues early''} (P241, L3). Although participants typically retained final oversight (Finding 3), verification was already embedded within these activities, reducing the perceived cost of AI action. Infrastructure \& Ops did not have statistically signigicant differences from Development. These findings \textbf{support H5}, indicating that accepted autonomy varies across SDLC categories, with significant differences observed for Meta-work, Design \& Planning, and Quality \& Risk.

\textbf{Robustness check}. We evaluated two alternative statistical operationalizations as robustness checks for our findings. First, a linear mixed-effects model reproduced the same directional patterns as the CLMM: AI Experience ($\beta = 0.12$, $p = 0.009$) and Risk Tolerance ($\beta = 0.16$, $p = 0.014$) were associated with higher accepted autonomy, whereas Identity was associated with lower accepted autonomy ($\beta=-0.07$, $p = 0.023$). SDLC-category effects were also preserved, with lower accepted autonomy for Design \& Planning ($\beta = -0.48$) and Meta-work ($\beta = -0.56$), and higher accepted autonomy for Quality \& Risk ($\beta = 0.19$), all $p < 0.05$.  Second, a binary logistic mixed model that dichotomizes responses at the decision-making boundary ($\le$L3 vs.\ $>$L3) reproduces the CLMM in direction and significance: Task Identity reduced the odds of accepting autonomy above L3 (OR = 0.85, $p = 0.018$), whereas AI Experience (OR = 1.29, $p = 0.010$) and Risk Tolerance (OR = 1.41, $p = 0.013$) increased those odds. All three SDLC-category contrasts were preserved: Design \& Planning (OR = 0.40), Meta-work (OR = 0.32), and Quality \& Risk (OR = 1.69), all $p{<}0.05$, relative to Development. Overall, the findings were robust to treating the outcome as ordinal, binary, or continuous. Full coefficients are reported in~\cite{supplemental}.

\subsection{RQ2: Crossing the Action and Decision-Making Boundaries}
\label{sec:rq2}

RQ1 identified the predictors of developers' overall accepted AI autonomy. In RQ2, we investigate whether the same predictors operate when AI gains authority to \emph{act} (the action boundary, L2$\to$L3) and \emph{decide} (the decision-making boundary, L3$\to$L4). To do so, we fit a GLMM at each of these transitions (Table~\ref{tab:transitions}) with the same predictors as RQ1.
 
\begin{table}[b]
\centering
\footnotesize
\setlength{\tabcolsep}{6pt}
\caption{\small Boundary-transition GLMMs: OR [95\% CI] at the action boundary (L2$\rightarrow$L3) and decision-making boundary (L3$\rightarrow$L4).}
\label{tab:transitions}
\vspace{-5px}
\begin{tabular}{@{}lcc@{}}
\toprule
\textbf{Predictor} &
\textbf{Action Boundary} &
\textbf{Decision-Making Boundary} \\
&
\textbf{(L2$\rightarrow$L3)} &
\textbf{(L3$\rightarrow$L4)} \\
\midrule

\multicolumn{3}{@{}l}{\textit{Task appraisals}} \\

Task Value
& 0.95 [0.79, 1.13]
& 1.05 [0.87, 1.27] \\

\textbf{Task Identity}
& 0.89 [0.74, 1.07]
& \textbf{0.78* [0.64, 0.95]} \\

\textbf{Task Accountability}
& \textbf{0.82* [0.70, 0.98]}
& 0.91 [0.76, 1.09] \\

\textbf{Task Demand}
& 0.97 [0.83, 1.14]
& \textbf{1.20* [1.05, 1.42]} \\

\addlinespace

\multicolumn{3}{@{}l}{\textit{Developer traits}} \\

SE Experience
& 0.92 [0.68, 1.24]
& 0.84 [0.62, 1.14] \\

\textbf{AI Experience}
& 1.27 [1.00, 1.61]
& \textbf{1.29* [1.01, 1.65]} \\

\textbf{Risk Tolerance}
& 1.40 [1.00, 1.96]
& \textbf{1.46* [1.03, 2.07]} \\

Technophilia
& 0.91 [0.67, 1.23]
& 1.06 [0.78, 1.45] \\

\midrule

$R^2_m, R^2_c$
& 0.103, 0.574
& 0.065, 0.543 \\

\bottomrule

\addlinespace[2pt]

\multicolumn{3}{@{}l}{\footnotesize $*\,p<.05$, $**\,p<.01$, $***\,p<.001$.}

\end{tabular}
\end{table}

\textbf{Finding 7: Accountability was negatively associated with the action boundary, lowering the odds of crossing it.} Task Accountability lowered the odds of crossing the L2$\rightarrow$L3 boundary (OR~=~0.82, $p{=}0.027$). Put simply, when participants felt highly accountable for a task, they were less willing to let AI produce work artifacts, even while remaining open to AI suggestions: \textit{``I do not want AI to `handle' any of it, given that I am accountable for it. But I welcome AI systems that provide feedback on my design and implementation''} (P174, L2). As P120 explained in the context of code review, \textit{``marking my approval puts my name on it and makes me partially responsible for it''} (P120, L2). This pattern is consistent with prior work showing that accountability increases vigilance and the desire to oversee outcomes before committing to them~\cite{tetlock1983accountability,lerner1999accounting}. The effect emerged precisely at the transition where AI outputs become reusable work artifacts and the consequences of automation errors become attributable to the individual, making automation complacency a more salient concern~\cite{parasuraman2010complacency}.

\textbf{Finding 8: At the decision-making boundary, Task Identity and Demands pulled in opposite directions.} By L3, participants had accepted AI as a collaborator capable of producing work artifacts under oversight. Crossing into L4 meant accepting AI's decisions by default with an optional veto, shifting the developer role from decision-maker to overseer. Task Identity was associated with lower odds of crossing this boundary (OR~=~0.78, $p{=}0.01$), whereas Task Demand was associated with higher odds (OR~=~1.20, $p{=}0.03$).

While accountability predicted participants' willingness to let AI act (Finding~7), the L3$\rightarrow$L4 boundary concerned who retained final judgment once AI had acted. Participants who viewed a task as central to who they were as professionals were more willing to accept AI-generated work than to relinquish the judgment surrounding it. As one participant explained, \textit{``I wouldn't want AI to handle final decision-making in high-stakes risk scenarios, especially where ethical judgment, human intuition, or deep domain context is critical [...] the responsibility should remain with experienced professionals''} (P10, L3). Prior work links identity to ownership, autonomy, and authorship over valued work outcomes~\cite{ryan2000self,kahn1990psychological,koestner2002attaining}; our findings suggest that this attachment is expressed most strongly at this boundary.

Task Demand exhibited the opposite pattern. Participants facing heavier demands were more willing to delegate decision-making, suggesting that the cost of exercising oversight can outweigh the value of retaining it. Consistent with job demands--resources theory~\cite{bakker2007job}, workload increased the appeal of transferring not only execution but also approval and coordination effort. One participant remarked, \textit{``As much as I love coding, there is simply too much work to do to want to keep doing it myself''} (P56, L5).
Together, these results suggest that crossing the decision-making boundary reflects a trade-off between preserving decision-making on identity-relevant work and offloading it when task demands become burdensome.

Overall, \textbf{H6 was partially supported}. Task Accountability was associated with lower odds of crossing the action boundary (L2 $\rightarrow$ L3), whereas Task Identity and Task Demands were associated with lower and higher odds, respectively, of crossing the decision-making boundary (L3 $\rightarrow$ L4).

\textbf{Finding 9: AI Experience and Risk Tolerance were positively associated with crossing the decision-making boundary.}
AI Experience (OR~=~1.29, $p{=}0.044$) and Risk Tolerance (OR~=~1.46, $p{=}0.034$) were associated with higher odds of crossing the L3$\rightarrow$L4 boundary. While Finding~8 showed that task appraisals predicted whether participants retained decision-making authority, these traits distinguish how readily participants were willing to grant it. For example, at one end, P276 saw little reason to withhold autonomy: \textit{``[\ldots] I have been using AI since the start of any incubating idea to shipping the product''} (P276, L5, Risk Tolerance~=~5). At the other end, more risk-averse participants held back: \textit{``The more technical the task is, the more I feel the need to have personal confidence in its correctness. I don't usually get that confidence when leaning heavily on [AI]''} (P368, L2, Risk Tolerance~=~1).

Overall, the action and decision-making boundaries answered to different appraisals: accountability to whether AI may produce the artifact, identity and demands to whether AI may decide without sign-off, with AI experience and risk tolerance easing that crossing.


\section{Cascading Locks of Meaningful Work Design}

Designing meaningful work means understanding how developers appraise it.
Individuals are meaning-makers; we seek significance and value in what we
do~\cite{lips2009discriminating}, and decades of work-design
research~\cite{humphrey2007integrating, bakker2007job} show that how we appraise
a task, along its relevance, identity, accountability, and demand, accounts for
much of the variance in work satisfaction and
productivity~\cite{humphrey2007integrating, lazarus1991emotion, roseman2001appraisal}.
Two of these appraisals hold the line on AI autonomy. 
A developer who feels \textit{accountable} for a task keeps AI at suggestion and produces
the artifact themselves (Finding 7). A developer for whom a task is central to their \textit{identity} does not relinquish their decision-making (Finding 8).

We picture this as a flight of cascading canal locks. The vessel is AI's autonomy on a
single task. (A developer may stay at L2 for a security-critical task, but cede autonomy at the L4 level for rote fault-isolation scans). The locks raise it one gate at a time. Accountability is the gate of the first lock, at the action boundary. Identity is the gate of the
second, at the decision-making boundary. 
The locks cascade because they run in order: the vessel reaches the second gate only after the first has let it
through, so a developer faces the identity decision only once accountability has
cleared AI to produce the artifact. 
The water that lifts the vessel is AI
fluency and risk tolerance. Both raise the autonomy developers accept and the
odds of clearing each gate (Findings 5, 9). Demand adds a
second inflow at the upper gate (Finding 8): a heavier workload makes developers readier to hand off decision-making to AI.

Viewed through the lens of cognitive appraisal, our findings point to anti-patterns (see Table~\ref{tab:antipatterns}) 
that appear when tool defaults set these boundaries.
Each theme below discusses how these appraisals can inform meaningful work design.

\begin{table*}[t]
\centering
\footnotesize
\setlength{\tabcolsep}{6pt}
\renewcommand{\arraystretch}{1.0}
\definecolor{rowshade}{gray}{0.92}
\caption{\small Design themes for meaningful work and the anti-patterns to protect against. }
\label{tab:antipatterns}
\vspace{-5px}
\begin{tabular}{@{}p{0.13\textwidth} l p{0.64\textwidth}@{}}
\toprule
\textbf{Theme} & \textbf{Anti-pattern} & \textbf{Description} \\
\midrule
\multirow{2}{0.16\textwidth}{\textbf{Accountability\\ retention}}
& \textbf{Deferred answerability} & Review steps are automated; responsibility offloaded until no one owns the result. \\
& \textbf{Right-shifted quality} & Quality gates migrate to verification; errors surface late and cost more to fix. \\
\midrule
\multirow{1}{0.16\textwidth}{\textbf{Demand\\ differentiation}}
& \textbf{Throughput stampede} & Incentives reward velocity; workload metric conflate toil and challenge as one quantity causing work shedding that once defined self-worth and identity. \\
\midrule
\multirow{4}{0.16\textwidth}{\textbf{Identity evolution}}
& \textbf{Shrinking fence} & Craft core defined by AI capability gaps, which shrinks with each release.\\
& \textbf{Hollow orchestrator} & Role title remains, but person cannot follow what they need to approve. Veto becomes nominal. \\
& \textbf{Expertise commoditization} & AI does the underlying work; the expertise that distinguished the role stops being scarce. \\
& \textbf{Thought homogenization} & Developers defer to the same AI for framing; judgment converges on the model shrinking independent thinking and innovation. \\
\midrule
\multirow{3}{0.16\textwidth}{\textbf{Identity formation}}
& \textbf{Severed pipeline} & Entry-level work automated to increase throughput; juniors lack scaffolding before facing hard problems.\\
& \textbf{Cognitive debt accrual}& Juniors offload learning work to AI; skill gaps hide until AI fails. \\
& \textbf{Mentoring-by-bot} & Onboarding automated; relationships needed to pass tacit knowledge never form. \\
\bottomrule
\end{tabular}
\vspace{-5px}
\end{table*}

\textbf{Accountability retention: keep answerability where the artifact is shaped.}
Most system-facing work is at L3 today: AI produces and a developer approves
before the work takes effect. At L4 only an optional veto remains, and
answerability moves from before the commit to after the failure. This is the
\emph{deferred answerability} anti-pattern: the developer sees the artifact when
verification reports a problem, not when it is built. The accountability
literature explains the cost. A developer who expects to answer for a decision
beforehand reasons carefully and self-critically; one who answers for it only
after the result is known defends the result
instead~\cite{lerner1999accounting, tetlock1983accountability}. Two design
choices can help. First, the L4 veto has to require real
engagement at the commit, so approval is more than just a ``rubber stamp''.
Second, developers should answer for how they reviewed the work, not only for the shipped product. Both keep answerability concrete and
attributable to a person~\cite{hall2017accountability}, and both belong in how
the role and workflow are built, and not delegated to a call for individual
diligence~\cite{ulloa2025product, frink1998toward}.

\textbf{Demand differentiation: separate toil from work that builds judgment}.
Demand was the one appraisal that pushed toward more AI autonomy at the
decision-making gate (Finding 8), and whether that push is safe depends on the type of demand. The job-demands literature distinguishes work that only
drains effort from work that drains it while building skill and mastery~\cite{bakker2007job, lepine2005meta}. Handing AI the repetitive, well-scoped work that consumes attention and teaches nothing costs a developer nothing. The cognitively hard work that
builds judgment is the work human oversight depends on.
A workload metric that counts only volume cannot tell the two apart, so under a
deadline it offloads both. This is the \emph{throughput stampede} anti-pattern. The design choice is to measure the two demands separately: route
toil to AI on purpose, and protect the hard, judgment-building work from the same
automation pressure rather than let a volume target decide its fate.

\textbf{Identity evolution: anchor the protected core in judgment.}
A developer's sense of craft can rest on one of two things: the work AI cannot
do yet, or the standard of judgment the role demands. The first is a moving
target. Each model release does part of what defined the core, the
\emph{shrinking fence}, until the title remains but its real work is gone, the
\emph{hollow orchestrator}. It fails in the labor market too: once AI does the
underlying work, any developer can direct it, and the expertise that once set a
developer apart is no longer scarce, the \emph{expertise commoditization} 
anti-pattern \cite{ehsan2026future}. The same deference to AI homogenizes a team. Finally, as developers lean on AI for
framing and first drafts, they start from the same suggestions and reason from
the same defaults, so their judgments converge on the model's, dissent thins, and
the team loses the range of independent thinking that open-ended work depends
on~\cite{anderson2022measuring}. 

Self-determination theory~\cite{ryan2000self} explains why an identity built on what AI cannot yet do is unstable. Ownership comes from work a developer values for its own sake, not from defending tasks against obsolescence, and once a tool closes the gap, that basis for identity is gone~\cite{ryan2000self, koestner2002attaining}. Identity economics adds the market half: an identity has value when it ties a person to a standard others recognize, and a standard of judgment persists even as capabilities become widely available~\cite{akerlof2000economics}.The design choice is to define the
core by the judgment the role requires and to write that into role definitions
and promotion criteria, where it is organizational rather than a personal bet on
what AI will not reach, and where developers reason from their own judgment
rather than defaulting to the model's.

\textbf{Identity formation: protect the work that trains judgment}.
Judgment is built by doing the work, which is embodied in the
entry-level work that seems most delegable to AI. Automate those tasks away and the skills needed to reach senior levels never form, the
\emph{severed pipeline}. Two failures feed it. Juniors who hand such work to AI are unaware of skill gaps, the \emph{cognitive
debt}~\cite{choudhuri2026thinking}; and a bot that manages onboarding delivers explicit knowledge but none of the tacit knowledge that only passes between
people, \emph{mentoring-by-bot}. The competence and ownership that ground a
developer's identity come from doing work pursued for its own
sake~\cite{ryan2000self, kahn1990psychological}, so the design choice is to keep a
share of entry work as practice rather than throughput. 
The developers who need to exercise judgment over AI-mediated systems learn it on exactly the work most tempting to automate away.

\section{Conclusion}
\label{sec:conclusion}

Our study of 448 developers found that the two boundaries around AI autonomy answer to different appraisals. Task accountability predicts whether AI may act and produce work artifacts. Task identity and demands predict whether it may decide on the developers' behalf. These appraisals reflect how developers relate to their work, so the accepted autonomy level can shift as AI capability, developer experience, and the work itself co-evolve. AI experience and risk tolerance already move it (Finding~5,~9), and the next model release may move it again. What is likely to hold steady underneath is the need for work that sustains meaning, competence, and identity.

A design response pinned to a list of tasks to withhold from AI is therefore brittle, whereas one grounded in meaningful work design endures. We offer the pairing of cognitive appraisals and our analysis instruments as a living framework that organizations can rerun as roles and tools change, to see where and why their developers draw these boundaries and whether adoption is breeding the anti-patterns in Table~\ref{tab:antipatterns}.

\textit{We have a choice} to set these boundaries deliberately, through how roles, workflows, and jobs are designed, or let tool defaults reset them with each release. The first path keeps the work meaningful as automation advances.

\vspace{-2mm}

\bibliographystyle{ACM-Reference-Format}
\bibliography{references}

\end{document}